\documentclass[twocolumn]{svjour3}         % twocolumn
\smartqed  % flush right qed marks, e.g. at end of proof
\usepackage[T1]{fontenc} % pour taper les lettres accentuées
\usepackage[latin1]{inputenc}
\usepackage{amsmath} % just math
\usepackage{amssymb} % allow blackboard bold (aka N,R,Q sets)
\usepackage{color}
\usepackage{caption}
\usepackage{bm}
\usepackage{ulem}
\usepackage{graphicx}
\begin{document}

\title{A slip model for micro/nano gas flows
induced by body forces}%\thanks{Grants or other notes
%about the article that should go on the front page should be
%placed here. General acknowledgments should be placed at the end of the article.}
%}
%\subtitle{Do you have a subtitle?\\ If so, write it here}
%\titlerunning{Short form of title}        % if too long for running head

\author{Q.D. To         \and
        C. Bercegeay \and
        G. Lauriat \and
        C. Léonard \and
        G. Bonnet \and
}

%\authorrunning{Short form of author list} % if too long for running head

\institute{Q.D. To  \at
              Université Paris-Est, Laboratoire Modelisation et
Simulation Multi Echelle, FRE 3160 CNRS, 5 Boulevard Descartes,
77454 Marne-la-Vall\'ee Cedex 2, France \\
              Tel.: +33-1-60957310\\
              Fax: +33-1-60957294\\
              \email{quy-dong.to@univ-paris-est.fr}           %  \\
%             \emph{Present address:} of F. Author  %  if needed
}

\date{Received: date / Accepted: date}
% The correct dates will be entered by the editor

\maketitle

\begin{abstract}
A slip model for gas flows in micro/nano channels induced by
external body forces is derived based on Maxwell's collision
theory between gas molecules and the wall. The model modifies the
relationship between slip velocity and velocity gradient at the
walls by introducing a new parameter in addition to the classic
Tangential Momentum Accommodation Coefficient. Three-dimensional
Molecular Dynamics simulations of helium gas flows under uniform
body force field between copper flat walls with different channel
height are used to valid the model and to determine this new
parameter. 
% \PACS{PACS code1 \and PACS code2 \and more}
% \subclass{MSC code1 \and MSC code2 \and more}
\end{abstract}
\begin{table}
\begin{center}
\caption*{NOMENCLATURE }
\begin{tabular}{l  l }
\hline
    $\lambda$ & Mean free path\\
    $H,B,L$ & Channel height, width and length\\
        Kn & Knudsen number\\
        $n,\rho$ & Number density, mass density\\
        $d$ & Molecular diameter\\
$L_s, L^*_s$ & Slip length, dimensionless slip length\\
$x,y,z$  & Cartesian coordinate\\
$\hat z$& Normalized coordinate\\
$\sigma_v$  & Tangential Momentum Accommodation Coefficient\\
$v,\hat v$ & Tangential velocity, normalized tangential velocity\\
$v_{slip},\hat v_{slip}$ & Slip velocity, normalized slip velocity\\
$ v_{ref}$ & Reference velocity\\
$N^\pm$ & Molecules going upward $N^+$ and downward $N^-$\\
& with respect to the control surface $s$\\
$N$ & Total number of molecules passing through $s$\\
$<v^\pm_x>$ & Average velocity of molecules going upward $<v_x^+>$ \\
& and downward $<v_x^->$ with respect to $s$\\
$v_s,v_w$ & Gas velocity near the wall, velocity of the wall\\
$<\tau>,\bar c$ & Mean collision time, thermal speed\\
$k_B, T$ & Boltzmann constant, absolute temperature\\
$m,\gamma_x$ & Molecular mass, acceleration along x-axis\\
$\alpha,\beta$ & Slip parameters of the present model\\
$v_\lambda$ & Gas velocity at distance $\lambda$ from the wall\\
$\mu,\mu^*,\bar\mu$ & Gas viscosity, scaled viscosity, kinetic theoretical\\
& viscosity\\
$V_i$ & Potential energy of atom $i$\\
$F,\rho_e,\phi$ & Potential functions of Embedded Atom Model \\
$\sigma_{a-b},\epsilon_{a-b}$ & Parameters of Lennard Jones potential between \\
&  molecules $a$ and $b$ \\
$r_{ij}$ & Distance between two molecules $i$ and $j$.\\
          \hline
\end{tabular}
\end{center}
\end{table}
\section{Introduction}
\label{sec-intro} The velocity of a fluid close  to a solid wall
is always different from the wall velocity even if the latter is
perfectly diffusive. Especially, when the channel height is
decreased to that of MEMS or NEMS devices (Micro/Nano
Electro-Mechanical Systems), this phenomenon becomes highly
important and must be taken into account. The Knudsen number
$\mathrm{Kn}$, the ratio between the mean free path $\lambda$ and
the characteristic length of the channel $H$, is the relevant
parameter to quantify the slip effects. The mean free path
$\lambda$ is usually defined as the average distance that
molecules travel between collisions and equal to
\begin{eqnarray}
&&\lambda  = \frac{1}{{\sqrt 2 n\pi d^2 }}
\end{eqnarray}
where $n$ is number density and $d$ is the effective molecular
diameter. According to Maxwell's model \cite{maxwell1879srg}, the
slip length $L_s$ in continuous fluid mechanics can be determined
via the Tangential Momentum Accommodation Coefficient, also
denoted by TMAC or $\sigma_v$ as follows
\begin{eqnarray}
&&L_s=\frac{2-\sigma_v}{\sigma_v}\lambda\label{eq-sliplength}
\end{eqnarray}
The slip velocity, $v_{slip}$ is calculated by the formula
\begin{eqnarray}
&&v_{slip}=L_s\left.\frac{\partial v}{\partial
z}\right|_w\label{eq-slipeqgen}
\end{eqnarray}
The term $\left.\frac{\partial v}{\partial z}\right|_w$ is the
normal derivatives of the tangential velocity component at the
walls, assuming that the normal to the wall is in the z-direction.
If the velocity and coordinate in Eq. (\ref{eq-slipeqgen}) are
scaled with a reference velocity $v_{ref}$ and the channel height
H, we have
\begin{eqnarray}
&&\hat{v}_{slip}=L^*_s\left.\frac{\partial \hat{v}}{\partial
\hat{z}}\right|_w,\quad L^*_s=\frac{L_s}{H}\label{eq-slipeqnom}
\end{eqnarray}
where $\hat{v}_{slip}=v_{slip}/v_{ref}$ and $\hat{z}=z/H$. The
term $L^*_s$ is called the dimensionless slip length and equal to
\begin{eqnarray}
&&L^*_s=\frac{2-\sigma_v}{\sigma_v}\mathrm{Kn}\label{eq-sliplengthnom}
\end{eqnarray}
when the Maxwell model is used. Consequently, for a given
accommodation coefficient, Eq. (\ref{eq-sliplengthnom}) predicts
that $L^*_s$ is proportional to Kn. When Kn tends towards zero, we
recover the no slip condition and when Kn increases, the slip
effect increases. The Maxwell model is widely used to describe the
slip at the walls because it only needs one parameter only,
$\sigma_v$. \\
\\
The physical meaning of $\sigma_v$ in
(\ref{eq-sliplength},\ref{eq-sliplengthnom}) is that if $M$
molecules arrive at the wall, $\sigma_vM$ of them are reflected
diffusively and the remaining $(1-\sigma_v)M$ molecules are
reflected specularly. Based on Eq. (\ref{eq-sliplength}), one can
determine TMAC by either experiments or Molecular Dynamics (MD)
simulations. Some experimentalists controlled macroscopic
quantities such as pressure and mass flow rate and made use of the
relationship with the slip velocity to find TMAC (see e.g.
\cite{arkilic2001mfa,colin2004vso}). Arkilic et al.
\cite{arkilic2001mfa} studied flows of nitrogen, argon and carbon
dioxide through rectangular silicon channels and found TMAC
ranging between 0.75 and 0.85. Colins et al. \cite{colin2004vso}
worked on the couples silicon-nitrogen and silicon-helium and
found a relatively high coefficient, 0.93. Recently, Maali and
Bhushan \cite{maali2008slm} studied confined air flow between a
spherical glass particle glued to the cantilever of an atomic
force microscopy (AFM) and a glass plate. They let the cantilever
oscillating and measured the hydrodynamic damping factor.
Consequently, a value of the slip length of about 118 nm was
obtained. It corresponds to an accommodation coefficient of 0.72.
Direct measurements of TMAC on the couple He-Cu by Seidl (see
\cite{finger2007mds,gadelhak1999fmm}) stated that the coefficient
depends on the collision angle between gas and solid, ranging from
0.6 to 1.0. On the other hand, Cao et al.
\cite{cao2004agm,cao2005tdt,cao2006esr} using MD approaches to
simulate flows, revealed that TMAC at Ar-Pt interface can be as
small as 0.2 and influenced by temperature and surface roughness.
Arya et al. \cite{Arya2003msk} simulated directly the collisions
between gas and wall. Their results showed the dependence of TMAC
on the wall's lattice structure. It remains constant as long as
the drift velocity is small enough (less than 100 m/s). Finger et
al. \cite{finger2007mds} used similar method as \cite{Arya2003msk}
to show that TMAC is affected by the adsorbed layer and their
results matched the previous experiment of Seidl on the couple
copper-helium. Generally speaking, both experimental and MD works
gave a rather scattering results of TMAC and reflected the
dependency of TMAC on many factors. Hence, this coefficient should
be understood as an effective one and to be used
with caution.\\
\\
Extensions of Maxwell's model to different configurations have
been considered in the past. When the Knudsen number increases,
Beskok (see \cite{karniadakis2005man,gadelhak1999fmm} and the
references cited therein) argued that higher order of the Knudsen
number and higher order derivatives of velocity must be used in
the slip equations. For curved surface, Lockerby et al.
\cite{lockerby2004vbc} accounted for the contribution of normal
velocity in the viscous stress. In all the aforementioned works,
the gas molecules are not subject to any external forces before
arriving at the wall, which is not applicable in the presence of
volume force field such as gravity, electrostatics, etc... With
these force fields, the slip velocity is different from flows
driven by pressure gradient and cannot be simply described by
(\ref{eq-sliplength}). In what follows, on the basis of kinetic
theory, we derive the new slip equation that accounts for the
external body force field.
\begin{figure}[h]
\begin{center}
\includegraphics[trim = 0cm 1.5cm 0cm 1.5cm, width=7cm]{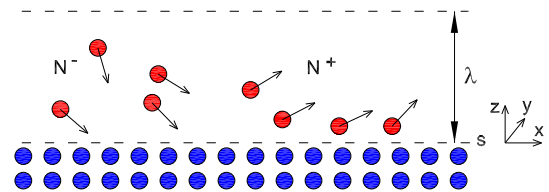}
\caption{Collision model between gas molecules and a solid wall.
$N^-$ and $N^+$ are respectively the number of molecules going
downward and upward and passing through the surface $s$ near the
wall within one unit time. If there is no accumulation of gas near
the wall, $N^-=N^+$. } \label{fig-scheme}
\end{center}
\end{figure}
\section{Slip model for micro/nano gas flows induced by body force}
\label{sec-slipmodel} In the following derivation, the gas flow is
assumed to be isothermal so that the influence of temperature on
the slip velocity is not taken into account. Using the similar
approach of Beskok (see \cite{karniadakis2005man}), let us
consider a control surface $s$ near and parallel to the wall (see
Fig. \ref{fig-scheme}). For a unit time, there are $N$ gas
molecules passing through the surface which are composed of $N^-$
molecules going downward and $N^+$ going upward with respectively
average tangential velocities, $<v_x^{-}>$ and $<v_x^{+}>$. The
gas average tangential velocity at the wall $v_s$ may be written
as
\begin{eqnarray}
&&N^{+}<v_x^{+}>+N^{-}<v_x^{-}>=Nv_s
\end{eqnarray}
The molecules that go upward are those that previously went
downward and were reflected at the surface $s$. Because the
reflection is either diffusive or specular by the fractions
$\sigma_v$ and $1-\sigma_v$, the following relation holds for a
wall moving at velocity $v_w$:
\begin{eqnarray}
&& N^{+}<v_x^{+}>=(1-\sigma_v)N^{-}<v_x^{-}>+\sigma_vN^{-}v_w
\end{eqnarray}
which is equivalent to
\begin{eqnarray}
&&
Nv_s=N^{-}\left[<v_x^{-}>+(1-\sigma_v)<v_x^{-}>+\sigma_vv_w\right]
\end{eqnarray}
Without external volume force, the $N^-$ molecules  colliding with
the wall come from one mean free path $\lambda$ away from the wall
without collision so that their velocity does not change $
<v_x^{-}>=v_\lambda$. We also assume  that $N^-=N^+=N/2$. It
follows
\begin{eqnarray}
&& 2v_s=\left[(2-\sigma_v)v_\lambda+\sigma_vv_w\right]
\end{eqnarray}
The Taylor development of $v_\lambda$ near the wall, $v_\lambda =
v_s + \lambda \left.\frac{\partial v}{\partial z}\right|_{w}$,
gives the relation for slip velocity $v_{slip}$ defined as
$v_s-v_w$ (first order Maxwell's relationship)
\begin{eqnarray}
&&v_{slip}=\alpha\left.\lambda \frac{\partial v}{\partial
z}\right|_{w}\quad \mathrm{with}\quad \alpha
=\frac{2-\sigma_v}{\sigma_v}
\end{eqnarray}
In the presence of uniform volume force, i.e. in the case where a
constant acceleration $\gamma_x$ is applied on each gas molecule,
$ <v_x^{-}>$ is no longer equal to $v_\lambda$. When impinging at
the wall surface, the term $\gamma_x \beta <\tau>$  should be
added to the average tangential velocity
\begin{eqnarray}
&& <v_x^{-}>=v_\lambda + \gamma_x \beta <\tau>. \label{v_x}
\end{eqnarray}
Above, $<\tau>$ is the mean time for the molecules to arrive at
the surface after the previous collision and is assumed to be
$\lambda/\bar c$ where $\bar c$ is the thermal speed of the gas.
For gases in local equilibrium, the thermal speed is estimated by
$\bar c=\sqrt{2k_BT/m}$. The constant $\beta$, introduced in the
last term of (\ref{v_x}), can be seen as a factor which accounts
for the differences between the idealized conditions used to
derive the slip model and the realistic ones.
In reality, these differences can be due to the following reasons \\
- the wall in the model is idealized as a surface and the gas wall
collisions only take place at this surface. In fact, the wall also
has an atomistic structure and the interaction force must be taken
into account at distance of several molecular diameters.\\
- after arriving at the wall, the gas molecules can stay near the wall during a finite duration of time before leaving. \\
\\
The value of $\beta$ is expected
to be close to unity. The slip equation with the corrected term
becomes
\begin{eqnarray}
&& v_{slip}=\alpha\lambda\left[\beta\frac{\gamma_x}{\bar c}+
\left. \frac{\partial v}{\partial z}\right|_{w}\right]
\label{eq-newslip}
\end{eqnarray}
Equation (\ref{eq-newslip}) is the new slip equation for general
flows in which the body force is involved. In what follows, we
study a particular case where the gas flow of viscosity $\mu$ is
induced by a constant body force $\rho\gamma_x$ along one
direction $x$. Without pressure gradient, the velocity profile is
given by
\begin{eqnarray}
&&  v(z)=
\frac{\rho\gamma_x}{2\mu}\left(\frac{H^2}{4}-z^2\right)+v_{slip},\label{eq-nsanalytical}
\end{eqnarray}
which after combined with the new slip model equation
(\ref{eq-newslip}) yields the dimensionless form
\begin{eqnarray}
&& \frac{\bar c\,v_{slip}}{\lambda\gamma_x}=\alpha\left[\beta+
\frac{1}{\mu^*\mathrm{Kn}}\right],\quad \mu^*=\frac{ \mu }{\bar
\mu}.\label{eq-modelslip}
\end{eqnarray}
In the above equation, $\mu^*$ is the scaled viscosity and $\bar
\mu $ is the kinetic theoretical viscosity
\cite{liou03mm,struchtrup2005mte} defined as $\bar
\mu=\frac{1}{2}\rho\lambda\bar c$. The dimensionless slip length
$L_s^*$ can be calculated accordingly
\begin{eqnarray}
&& L_s^*=\alpha\mathrm{Kn}\left[\beta\mu^*\mathrm{Kn}+
1\right]\quad \mathrm{with}\quad L_s^*=L_s/H
\end{eqnarray}
We can deduce that the derived slip length is second-order
dependent of the Knudsen number. The influence of the volume force
on the slip length becomes thus important when $\mathrm{Kn}$ is
large enough. The ratio $L_s^*/\mathrm{Kn}$ is no longer equal to
the constant $\alpha$ but is dependent on the channel
characteristic dimension $H$ via a composite parameter
$(\mu^*\mathrm{Kn})^{-1}$. If $\mathrm{Kn}$ is increased and the
variation of fluid viscosity $\mu$ is small, we shall observe an
increase in the ratio $L_s^*/\mathrm{Kn}$. In what follows, we use
the molecular dynamics approach to valid this prediction and
determine as well the two new model parameters $\alpha$ and
$\beta$.
\begin{figure}[h]
\begin{center}
\includegraphics[trim = 0cm 0cm 0cm 0cm, clip, width=10cm]{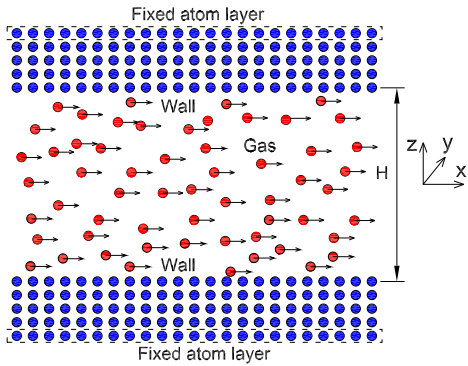}
\caption{One dimensional flow induced by external volume force.
The simulation domain is periodic in directions $x,y$. The
external volume force in direction $x$ is produced by applying
additional constant acceleration $\gamma_x$ on each gas atoms.}
\label{fig-flowscheme}
\end{center}
\end{figure}
\section{Molecular dynamics validation}
\label{sec-moldynval} 
\begin{figure}[h]
\begin{center}
\includegraphics[ width=8cm]{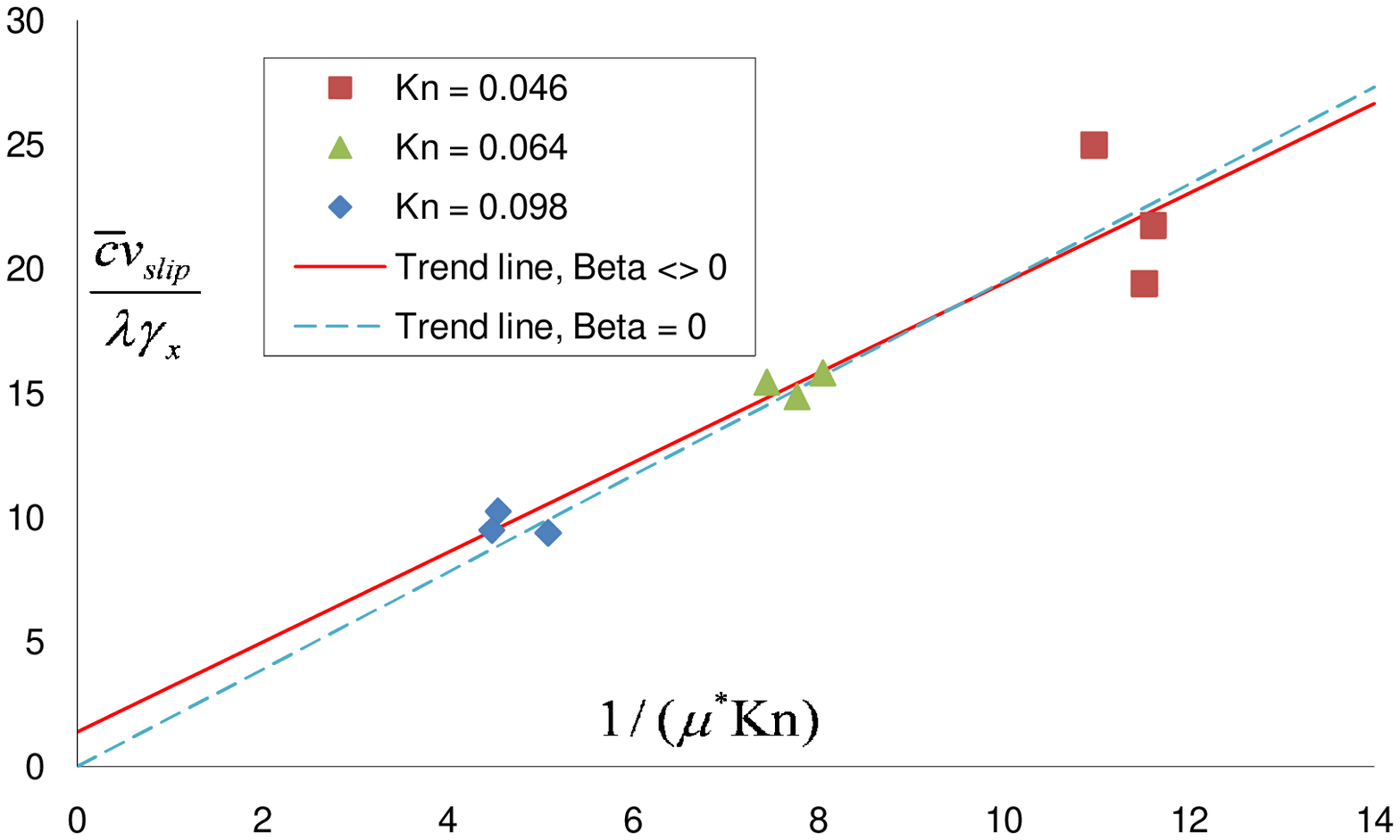}
\caption{Relations between two dimensionless quantities
$\frac{\bar c\,v_{slip}}{\lambda\gamma_x}$ and $1/(\mathrm{Kn}
\mu^*)$. The solid line represents the equation
(\ref{eq-modelslip}) which fits the numerical results. This line
makes an angle $\arctan\alpha$ with the horizontal axis and cuts
the vertical axis at $\alpha\beta$. The dashed line represents
(\ref{eq-modelslip}) with $\beta=0$ that fits the numerical
results.} \label{fig-slipresults}
\end{center}
\end{figure}

In order to recover the dependence of slip
velocities on the Knudsen number and volume force, we simulate a
flow induced by uniform external volume force. The gas/solid
couple investigated is helium (He) and copper (Cu). The choice of
the couple He-Cu is motivated by two reasons: firstly, the
accommodation coefficients derived experimentally and numerically
are shown to be in good agreement (see for example ref.
\cite{finger2007mds}) and, secondly the interaction potentials
between these atomic species have been widely used (see ref.
\cite{foiles86eam,daw1984eam,plimpton1995fpa,finger2007mds,allen1989csl}).
However, the present method can be applied to any gas/wall system
as long as the interaction potential between the atoms is properly
provided. Because the slip length depends on what happens at the
interface,
the final results are strongly affected by the choice of the system. \\
\\
The channel is made of two parallel solid walls, each of them
contains 9 layers (or 4 lattice units) of Cu atoms placed at fcc
lattice sites. The lattice constant is chosen initially equal to
3.615 \AA. The interaction forces between Cu atoms are derived
from the EAM potential \cite{foiles86eam,daw1984eam}
\begin{eqnarray}
&& V_i =  F\left[\sum_j \rho_e (r_{ij})\right]  +
\frac{1}{2}\sum_{i \ne j} \phi (r_{ij})
\end{eqnarray}
where $V_i$ is the potential energy of atom $i$ composed of the
binary potential $\phi(r_{ij})$ and embedded potential $F$
accounting for electron density contribution $ \rho_e$. The
interaction forces for the couples He-He and He-Cu are derived
from the
 widely used Lennard-Jones potential
\begin{eqnarray}
&&V_i = \sum_j4\varepsilon \left[ \left( {\frac{\sigma }{r_{ij}}}
\right)^{12}-\left( \frac{\sigma }{r_{ij}} \right)^6 \right]
\end{eqnarray}
The parameters used in this paper are those from
\cite{allen1989csl,finger2007mds}, $\varepsilon_{He-He}=0.00088$
eV, $\sigma_{He-He}=2.28$ \AA, $\varepsilon_{He-Cu}=0.0225$ eV,
$\sigma_{He-Cu}=2.29$ \AA. The He-Cu parameters have been
calculated using the Lorentz-Berthelot mixing laws from the He-He
and Cu-Cu parameters in \cite{allen1989csl}. The last atom layer
at the two walls is fixed (see Fig. \ref{fig-flowscheme}). The
remaining atoms of the wall and the gas are maintained at the same
temperature $120$ K by a usual
scaling method after removal of the mean velocity. On average, the wall is kept immobile, i.e $v_w=0$.\\
\\
In this work, the global number density of the gas are kept
constant $n=0.0026$ \AA$^{-3}$  (or $\rho=1.78\times 10^{-14}$
pg/\AA$^{3}$) while the channel height $H$ and acceleration
$\gamma_x$ is varied to obtain results for different global
Knudsen numbers and volume forces $f$. At the small density number
and temperature of our simulations (e.g. 120 K), the helium is in
gaseous phase. Three values of $H=361$, 260 and 174 \AA \,
corresponding respectively to Kn=0.046, 0.064, 0.098 are
considered. These three Kn numbers are chosen to fall into the
range $(0.01,0.1)$ in order to assure the slip flow regimes in our
test cases according to \cite{karniadakis2005man}. The two other
dimensions of the simulation box along x-axis (the flow direction)
and along y-axis denoted shortly by length $L$ and width $B$ are
$L=H$ and $B=H/2$. For each channel height, the acceleration
applied on each atom is varied as $\gamma_x=0.0036$, 0.012, 0.024
\AA/ps$^2$. The computations are carried out by using LAMMPS, an
open source parallelized code \cite{plimpton1995fpa} on an IBM
Power6 machine. The equations of the particle motion are
integrated using a Leapfrog-Verlet algorithm with a time step
$0.002$ ps. The steady state is achieved after $5\times10^6$ time
steps and it takes another $2\times10^6$ time steps for the
average process. All the models are constructed in 3D with the
parameters given in Table \ref{tab-modelparameters}. Due to the
high density of the solid wall, an important number of Cu atoms
are considered. The largest model, case Kn=0.046, involves 289 500
molecules and takes 8 hours of computation on 512 processors.
\begin{table}
\begin{center}
\caption{Size of the simulation box and number of helium and
copper atoms for different Knudsen numbers}
\begin{tabular}{c  c  c  c  c  c }
\hline
    Kn & H [\AA]& L [\AA] & B [\AA] & Cu atoms & He atoms  \\
    \hline
    0.046 & 361 & 361 & 180 & 230 700 &58 800 \\

    0.064 & 260 & 260 & 130 & 51 840 &22 680
 \\

    0.098 & 174 & 174 & 87 & 23 040 &6 624
 \\
          \hline
\end{tabular}
\label{tab-modelparameters}
\end{center}
\end{table}
\\
\begin{figure}[h]
\begin{center}
\includegraphics[ width=9cm]{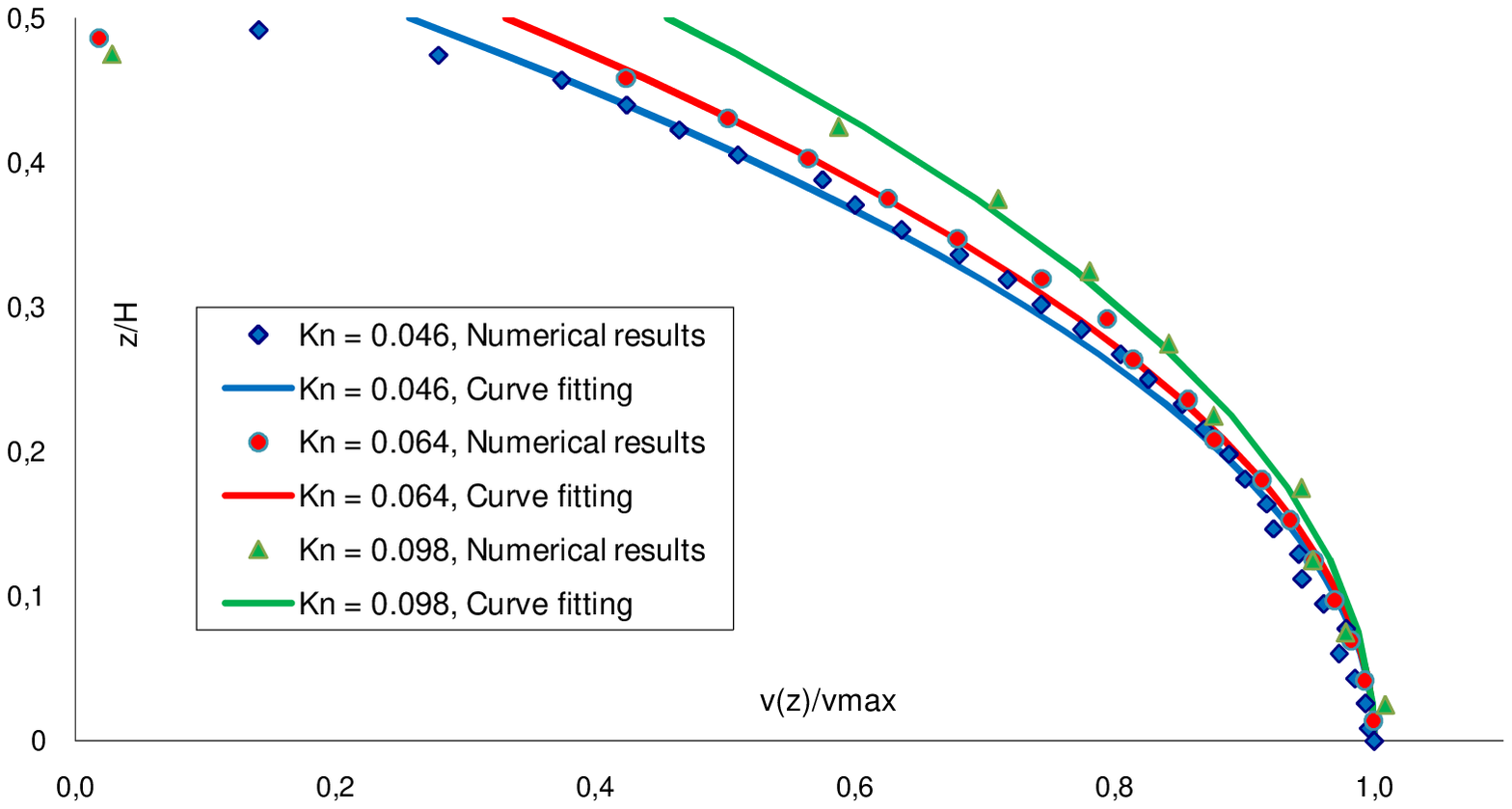}
\caption{Velocity profiles for the upper half of the channel
obtained by MD simulation (marker line) and fitted by analytical
formula (solid line). In the figure, the acceleration $\gamma_x$
is kept constant at $0.012$ \AA/ps$^2$ while the Knudsen number is
varied. The vertical and horizontal axis represent normalized
coordinate $z/H$ and velocity $v(z)/v_{\max}$ along the flow
direction. The velocity $v_{\max}$ is the maximal velocity found
in the middle of the channel.} \label{fig-velocityprofile}
\end{center}
\end{figure}
\\
In order to obtain accurate solutions, the channel has been
divided in a sufficiently large number of layers for the average
procedure. For $\gamma_x=0.012$ \AA/ps$^2$ and different Knudsen
numbers, the velocity profiles in the upper half channel are
plotted in Fig. \ref{fig-velocityprofile}. We find good agreement
between analytical solution (Eq. \ref{eq-nsanalytical}) and the
numerical results in the major part of the channel. The global
viscosity of the fluid $\mu$ and slip velocity $v_{slip}$ are
obtained by fitting the numerical velocity profiles (Fig.
\ref{fig-velocityprofile}) with equation (\ref{eq-nsanalytical}),
given in columns 3, 4 of Table \ref{tab-numresults}. It is noted
that the determination of $\mu$ is based the curvature of the
velocity profile and the average density $\rho$, regardless of the
redistribution of density at the steady state.
\begin{table}
\begin{center}
\caption{Numerical results from MD simulations}
\begin{tabular}{c  c  c  c  c  c  c}
\hline
    Kn & $\gamma_x$  & $\mu\times 10^{5}$ & $v_{slip}$  & $\frac{L_s^*}{\mathrm{Kn}}$ & $\frac{1}{\mu^*\mathrm{Kn}}$ & $\frac{\bar c\,v_{slip}}{\lambda\gamma_x}$\\
    & [\AA/ps$^2$] & [Pa.s] & [\AA/ps] & & & \\
    \hline
    0.046 & 0.0036 & 1.89 & 0.18 & 1.87 &11.60 & 21.80\\
          & 0.012 & 1.91 & 0.54 & 1.69 &11.50 & 19.40\\
          & 0.024 & 2.00 & 1.38 & 2.20 & 11.30 & 25.00\\
          \hline
    0.064 & 0.0036 & 1.95 & 0.14 & 1.97 &8.05 & 15.80\\
          & 0.012 & 2.02 & 0.43 & 1.91 &7.77 & 14.90\\
          & 0.024 & 2.11 & 0.89 & 2.10 &7.44 & 15.50\\
          \hline
    0.098 & 0.0036 & 2.03 & 0.08 & 1.90 &5.08 & 9.41\\
          & 0.012 & 2.31 & 0.27 & 2.12 &4.48 & 9.51\\
          & 0.024 & 2.27 & 1.38 & 2.26 &4.54 & 10.03\\
          \hline
\end{tabular}
\label{tab-numresults}
\end{center}
\end{table}
\\
\\
From the results reported in Table \ref{tab-numresults}, it can be
seen that $L_s^*/\mathrm{Kn}$ tends to increase with
$\mathrm{Kn}$. This trend cannot be accounted for by the usual one
parameter model. In order to determine the two constants $\alpha$
and $\beta$ in our proposed model (see Eq. \ref{eq-modelslip}), we
plotted two dimensionless quantities $\frac{\bar
c\,v_{slip}}{\lambda\gamma_x}$ and $(\mathrm{Kn}\mu^*)^{-1}$ from
the last two columns of Table \ref{tab-numresults} into Fig.
\ref{fig-slipresults}. We  found the straight line that best fits
these data. In the framework of our problem, the two values
$\alpha=1.81$ and  $\beta = 0.76$ were determined. As expected,
the value $\beta$ is of order unity while $\alpha$ corresponds to
an accommodation coefficient $\sigma_v=0.71$ which is in the
experimental range [$0.66$, $1.0$] by Seidl, reported in the paper
of \cite{finger2007mds} for the couple He-Cu. It is noted that
these two bound limits $0.66$ and $1.0$ correspond to the two
extreme impinging at angles $70°$ and $10°$ of helium molecules at
a copper wall in Seidl's experiments (see Tab. \ref{tab-Seidl}).
\begin{table}
\begin{center}
\caption{Seidl experimental data on the couple He-Cu (see
\cite{finger2007mds}) where TMAC depends on the collision angle of
the gas molecules }
\begin{tabular}{c  c  c  c  }
\hline
    Angle & Seidl & Seidl lower & Seidl  upper \\
& variability  &  TMAC value &  TMAC value \\
    \hline
    10 & 0.100 & 0.86 & 1.06 \\
    20 & 0.080 & 0.81 & 0.97 \\
    30 & 0.065 & 0.77 & 0.90\\
     40 & 0.050 & 0.73 & 0.83 \\
     50 & 0.040 & 0.71 & 0.79\\
      60 & 0.030 & 0.69 & 0.75\\
       70 & 0.020 & 0.66 & 0.70 \\
          \hline
\end{tabular}
\label{tab-Seidl}
\end{center}
\end{table}
\\
\\
The impact of parameter $\beta$ on slip velocity depends on the
relative importance between the two terms $\beta$ and
$(\mathrm{Kn}\mu^*)^{-1}$. For large Kn, the role of coefficient
$\beta$ becomes important, e.g. up to $20\%$ when Kn = 0.098 while
for small Kn, e.g. Kn = 0.046, this effect is almost negligible
($6\%$). In order to compare with the classical model that
involves a single parameter $\alpha$, the numerical results are
also fitted with a line passing through origin, the dashed line in
Fig. \ref{fig-slipresults}. The coefficient $\alpha$ predicted by
the classical model takes the higher value $\alpha=1.95$ and shows
more discrepancies with respect to the simulation results.
\section{Concluding remarks and discussion}
A new slip model for flows with volume force has been introduced.
Two parameters with physical meanings that relates slip velocity
and velocity gradient at walls are suggested. The first one is the
traditional tangential momentum accommodation coefficient, as in
the widely used Maxwell model, while the second one accounts for
the additional velocity that a molecule encompasses owing to the
applied force before striking at a the solid surface. Molecular
dynamics calculations applied on the He-Cu couple allowed to
determine both TMAC and the new parameter $\beta$ introduced in
the present work. TMAC is found to be in good agreement with
experimental data. In the $v_{slip}$ expression, the effect of
parameter $\beta$ increases with $\mathrm{Kn}$, i.e. when the
limit of transitional flow is reached. \\
\\
The present model is useful for the global analysis of flows
without knowing the presence of the Knudsen layer. In our MD
simulations, both higher density and slower motion of gas
molecules are observed near the walls and cause deviation of the
numerical velocity from the Navier-Stokes solution. Similar
phenomenon was encountered and discussed in \cite{cao2005tdt} from
molecular viewpoint. Generally, when a molecule reaches a surface,
it does not bounce back immediately but is often trapped by the
potential well. The molecules remains near the wall for some time,
interacts with many other solid atoms before escaping. Accounting
for the Knudsen layer should leads to more accurate results (see
Lockerby et al. \cite{lockerby2005uhoc,lockerby2008modelling}). A
model involving
body force as well as Knudsen layer may be expected in the near future. \\
\begin{acknowledgements}
The authors acknowledge the French National Institute for Advances
in Scientific Computations (IDRIS) for computational support of
this project through grant No.i2009092205. We also wish to thank
the reviewers for the given comments that help to improve the
quality of this paper.
\end{acknowledgements}

% BibTeX users please use one of
\bibliographystyle{spbasic}      % basic style, author-year citations
%\bibliographystyle{spmpsci}      % mathematics and physical sciences
%\bibliographystyle{spphys}       % APS-like style for physics
%\bibliography{}   % name your BibTeX data base

% Non-BibTeX users please use

\end{document}